\documentclass[a4paper,prb,twocolumn,amsmath,amsfonts,amssymb,superscriptaddress,nofootinbib,showpacs]{revtex4}

\usepackage{graphicx}
\usepackage{amsmath}
\usepackage{float}
\usepackage{longtable}
\usepackage{epsfig}
\usepackage{amssymb}
\usepackage{amsfonts}
\usepackage{latexsym}
\usepackage{theorem}
\usepackage{bm}
\usepackage{psfrag}
\usepackage[dvips]{color}
\usepackage[cmtip,arrow]{xy}

\newtheorem{thm}{Theorem}
\newtheorem{defi}[thm]{Definiton}
\newtheorem{lem}[thm]{Lemma}

\newenvironment{proof}[1][{}]{\par\vskip12ptplus30pt\noindent{\it Proof{#1}:\ }}%
{\hfill$\blacksquare$\par\vskip12ptplus60pt}

\def\1#1{{\bf #1}}
\def\2#1{{\mathcal #1}}
\def\4#1{{\tt #1}}
\def\5#1{{\sf #1}}
\def\6#1{{\mathfrak #1}}
\def\7#1{{\Bbb #1}}
\def\8#1{{\rm #1}}
\def\9#1{{\mathcurl #1}}

\DeclareFontFamily{OT1}{rsfs}{}
\DeclareFontShape{OT1}{rsfs}{m}{n}{<-7> rsfs5 <7-10> rsfs7 <10-> rsfs10}{}
\DeclareMathAlphabet\mathcurl{OT1}{rsfs}{m}{n}

\def\clb{\color{blue}}

\definecolor{grey}{rgb}{0.5,0.5,0.5}

\def\B{{\boldsymbol B}}

\def\I{\openone}

\def\tr{\8{tr}}
\def\adj{\dagger}
\def\U{\boldsymbol U}
\def\ket#1{|#1\rangle}
\def\bra#1{\langle#1|}

\newcommand{\sprod}[2]{\langle{#1},{#2}\rangle}
\newcommand{\expec}[1]{\langle{#1}\rangle}
\def\revddots{\mathinner{\mkern1mu\raise\p@
	\vbox{\kern7\p@\hbox{.}}\mkern2mu
	\raise4\p@\hbox{.}\mkern2mu\raise7\p@\hbox{.}\mkern1mu}}

\begin{document}

\title{Maximally entangled fermions}
\author{Dirk-M. Schlingemann}
\email{d.schlingemann@tu-bs.de}
\affiliation{Institut f\"ur
Mathematische Physik, Technische Universit\"at Braunschweig,
Mendelssohnstra{\ss}e~3, 38106 Braunschweig, Germany}
\affiliation{ISI Foundation, Quantum Information Theory Unit,\\
 Viale S. Severo 65, 10133 Torino, Italy}
\author{Marco Cozzini}
\email{cozzini@isi.it}
\affiliation{Dipartimento di Fisica, Politecnico di Torino,\\
Corso Duca degli Abruzzi 24, I-10129 Torino, Italy}
\affiliation{ISI Foundation, Quantum Information Theory Unit,\\
 Viale S. Severo 65, 10133 Torino, Italy}
\author{Michael Keyl}
\email{m.keyl@tu-bs.de}
\affiliation{Institut f\"ur
Mathematische Physik, Technische Universit\"at Braunschweig,
Mendelssohnstra{\ss}e~3, 38106 Braunschweig, Germany}
\affiliation{ISI Foundation, Quantum Information Theory Unit,\\
 Viale S. Severo 65, 10133 Torino, Italy}
\author{Lorenzo Campos Venuti}
\email{campos@isi.it}
\affiliation{ISI Foundation, Quantum Information Theory Unit,\\
 Viale S. Severo 65, 10133 Torino, Italy}
\date{\today}

\begin{abstract}
Fermions play an essential role in many areas of quantum physics and it is
desirable to understand the nature of entanglement within systems that consists
of fermions. Whereas the issue of separability for bipartite fermions has
extensively been studied in the present literature, this paper is concerned with
maximally entangled fermions. A complete characterization of maximally entangled
quasifree (Gaussian) fermionic states is given in terms of the covariance
matrix. This result can be seen as a step towards distillation protocols for
maximally entangled fermions.
\end{abstract}
\maketitle
\section{Introduction}
\label{sec-intro}

Since fermions play an essential role in almost all areas of quantum physics,
it is natural to study the nature of entanglement within fermionic systems. The notion of entanglement relies on the bipartite or even
multipartite structure of the underlying quantum system. We focus here on the
bipartite case by considering two parties, called Alice and Bob. To determine a
bipartite system, one has to specify the observables that are accessible within
Alice's and Bob's laboratories. The collection of all possible measurements that
can be made is described by the {\em global observable algebra}. Following the
reasoning of Ref.~\cite{ZarLidLL03}, a bipartite system can mathematically be
described by a pair of {\em local observable subalgebras} sitting inside the
global observable algebra. The two local algebras represent the observations that
can be made within Alice's and Bob's laboratories, respectively. The essential requirement is that each measurement performed by Bob cannot disturb the measurements made within
Alice's laboratory. Therefore, each of Alice's operators has to commute with all
of Bob's operators. Once a bipartite system is fixed by choosing the local
algebras, the notion of entanglement can be introduced {\em relatively to this
choice}: product states are characterized by the property that the expectation
values of a product of one of Alice's operators and one of Bob's operators
factorize. Taking the convex hull of all product states yields the set of
separable states and the issue of entanglement is to detect those states which
are not separable. We refer the reader to Refs.~\cite{BanCirWol07, Mor04}, where
the characterization of separable states on fermion systems is discussed.

In comparison to the considerations in Ref.~\cite{BanCirWol07}, we focus here on the
"converse" of separability, namely the characterization of maximally
entangled fermions. When dealing with fermions, the main issue is that, due to
the canonical aniticommutation relations (CAR) of Fermi fields, generic operators of Alice and Bob do not commute among each other. As we shall see, the standard bipartite structure discussed above can be restored either by restricting the analysis to even products of Fermi fields, as mainly done in this paper, or by introducing a twisted tensor product structure which automatically accounts for the alternating signs related to the commutation of Fermi fields belonging to different subsystems.

The restriction to even products of Fermi fields corresponds to the requirement
that physical observables preserve the parity of the particle number.
Since correlation experiments for the detection of entanglement are
built from observables that belong to the local subsystems of Alice and Bob, it is natural to restrict the analysis to operators which preserve not only the global parity, but also the local parity. In mathematical terms, we will then exploit the
restriction to the even-even part of the global algebra.
Consequently, we shall speak of {\em accessible} entanglement, relative to the restriction under consideration.
Clearly, once the even-even part of the algebra is considered, operators belonging to different subsystems commute among each other, giving rise to a tensor product structure analog to the qubit case.

As far as this restriction is concerned, our main result is the full characterization of quasifree (i.e., Gaussian) maximally entangled states.
We prove that, in order to be maximally entangled, a fermionic quasifree state must have a covariance matrix of a well-defined form, which is unique up to local Bogolubov transformations.

This analysis obviously requires a precise definition of fermionic maximally entangled states in the presence of the above restriction.
To this purpose, we exploit the conservation of local parity to further decompose the operator algebra.
Indeed, it turns out that the structure of Alice's and Bob's observable algebras
corresponds to the following situation: Alice possesses a direct sum of matrix
algebras of the form $\bigoplus_x \6A_x\otimes\I_x$ whereas Bob possesses the
algebra $\bigoplus_x \I_x\otimes \6B_x$, where
$\6A_x=\6B_x=M_d(\7C)$ are full matrix algebras.
Here, the label $x$ refers to the number parity of the two
subspaces (four combinations are possible, i.e., $++,+-,-+,--$, where $+$ and
$-$ correspond to even and odd, respectively), and, assuming that both Alice and
Bob own $n$ fermionic modes, one has $d=2^{n-1}$.
The direct sum structure of the observable algebras can be interpreted as a {\em
super-selection rule}, where the label $x$ plays the role of a conserved charge
or quantum number. Therefore, we are faced here with the issue of entanglement
in the presence of a super-selection rule.

While correlation experiments for the detection of entanglement only involve observables
belonging to the quasi-local
observable algebra given by $\bigoplus_x \6A_x\otimes\6B_x$, the fermionic state $\varrho$ under investigation can be prepared in a more general way and can be thought of as belonging to the global observable algebra.
\footnote{Differently from the quasi-local observable algebra, which preserves local parity, the global observable algebra only preserves global parity and therefore includes operators
which cannot be expressed as products of local observables.}
The restriction of $\varrho$ to the even-even part of the algebra, i.e., to the quasi-local observable algebra, is given by a density operator $\varrho^{++}$
of block-diagonal
form, $\varrho^{++}=\bigoplus_x p_x \varrho_x$, where $p_x$ is a normalized probability
distribution and $\varrho_x$ is a density matrix in $\6A_x\otimes\6B_x$. Pure
states are given by rank-one projections $\sigma_x$ in $\6A_x\otimes\6B_x$ and
the situation is effectively the same as for a simple tensor product since there
is only one non-zero term that contributes to the direct sum. A suitable
entanglement measure is the entropy of entanglement
$E(\sigma_x)=S(\tr_{\6B_x}(\sigma_x))$ where $\tr_{\6B_x}(\sigma_x)$ is the
reduced density matrix on Alice's subblock $\6A_x$ and
$S(\rho)=-\tr[\rho\log_2(\rho)]$ is the von Neumann entropy. If we consider a
density operator of the form $\varrho^{++}=\bigoplus_x p_x \sigma_x$ with rank-one
projections $\sigma_x$, then (due to the superselection rule) this is the {\em
unique} convex decomposition of $\varrho^{++}$ into pure states. For this class of
states the {\em entanglement of formation} $E_F$ can easily be computed. It is
just given by the mean value of the entropy of entanglement of each block, i.e.,
$E_F(\varrho^{++})=\sum_x p_x E(\sigma_x)$. Now we can ask for those states which
maximize the entanglement of formation. They are given by all density operators
of the form $\varrho^{++}=\bigoplus_x p_x \sigma_x$ where $\sigma_x$ is a projection
onto a maximally entangled vector. For all these states we find for the
entanglement of formation the maximal value $E_F(\varrho^{++})=\log_2(d)$. This is
precisely the notion of maximal entanglement that we apply here to the
investigation of entanglement in fermionic systems.

The discussion of fermionic entanglement can also be applied to
investigate entanglement in spin chain systems with the help of the
Jordan-Wigner transformation. Then, preservation of global parity corresponds
to preservation of total magnetization. Since
magnetization does not give rise to superselection rules, a fermionic system
formally arising from the Jordan-Wigner transformation of a spin chain is not
necessarily subject to the physical constraints discussed above. In particular,
the two local observable algebras are not restricted to the even part, given by
even products of Fermi fields, and their commutativity is not guaranteed. In such a context,
it is useful to consider the {\em twisted} tensor product structure mentioned above, which
takes into account the anticommutation relations of Fermi fields
similarly to the Jordan-Wigner transformation.

\begin{figure}
 \centering
 \includegraphics[width=9cm]{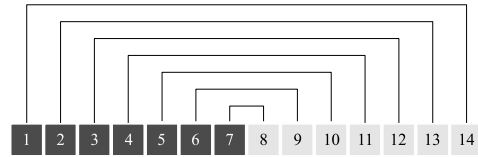}
 \label{fig:01}
\caption{The figure illustrates a spin chain that consists of two blocks of
spins. Alice controls the block $\{1,2,3,\cdots, 7\}$ (dark grey squares) and
Bob possesses the block $\{8,9,\cdots, 14\}$ (light grey squares). The black
lines are connecting those positions which are prepared in a maximally
entangled two qubit state.} 
\end{figure}

The outline of the paper is the following. We start by giving an elementary
overview (Section~\ref{overview}) of the basic ideas, and discuss, as an
instructive example, the simplest non-trivial case which corresponds to two
fermion modes per part. We state our main result on the characterization
of quasifree maximally entangled fermions in terms of this example first.
The structure of general bipartite fermion systems in finite (but arbitrary
large) dimensions is presented in Section~\ref{sec-1}. We take here advantage of
Araki's self dual formalism \cite{Ara70,Ara87}, which enables to write many
formulas in a more handy form.
The main result on the characterization of maximally entangled quasifree
fermion states is derived in Section~\ref{sec-2}. We give here an explicit
formula for the covariance matrix for a maximally entangled quasifree fermion
state. We also show that, up to local Bogolubov transformations, there is a
unique maximally entangled quasifree fermion state. Bogolubov transformations
are reversible physical operations that are characterized by the property to
map quasifree states into quasifree states (see Refs.
\cite{Ara70,Ara87,DirFanPog07} for general quasifree completely positive maps).
We also discuss a one-parameter group of (global) Bogolubov
transformations (which can be interpreted as the unitary time evolution due to a
proper interaction) mapping a product state into a maximally entangled fermion
state for a suitable value of the associated parameter (time).
As a further example, we consider in Section~\ref{sec-3} a spin chain
system consisting of two equally sized blocks of spins. The system is prepared
in a state where each site of one block is maximally entangled with a companion
site of the other block, according to Fig.~\ref{fig:01}. Applying the
Jordan-Wigner transformation the corresponding state on the fermion algebra is
quasifree and maximally entangled in the sense of our basic definition.
Finally, in Sec.~\ref{sec-twisted} we provide the description of the twisted tensor product structure which is necessary to extend the discussion beyond the even-even subalgebra.

\section{Overview}
\label{overview}
In order to have a closer look at the structure of Alice's and Bob's observable
algebras, we consider first the most simple example which consisits  of two
fermionic modes per part. The creation and annihilation operators $c_i^\dagger,c_i$ (generically denoted by $c^\#_{i}$ in the following)
are given here in the ``occupation number representation''. By fixing an
ordering of the modes first, this representation is given on the
$2^4=16$-dimensional Hilbert space $\2H$ that is spanned by basis vectors
$\ket{q_{1}q_{2} q_{3} q_{4}} = (c^\adj_1)^{q_1}(c^\adj_2)^{q_2}(c^\adj_3)^{q_3}(c^\adj_4)^{q_4}\ket{0000}$, labeled by binary strings of length four, i.e., $q_i=0,1$.
The 16-dimensional Hilbert space $\2H$ can be
decomposed into a direct sum of four 4-dimensional subspaces. Each vector
$\Psi\in\2H$ can be represented by a block decomposition
\begin{equation}\label{block-decomp-intro}
\Psi=
\left(\begin{array}{c}
\Psi_{++}\\
\Psi_{+-}\\
\Psi_{-+}\\
\Psi_{--}
\end{array}\right) \ ,
\end{equation}
where the four block vectors $\Psi_{xy}\in\7C^2\otimes\7C^2$, with
$x,y=\pm$, are simultaneous eigenvectors of the parity operators of both
subsystems. For example, $\Psi_{++}\in\2H_{++}={\rm
span}\{\ket{0000},\ket{1100},\ket{0011},\ket{1111}\}$.
Alice's observable algebra is generated by the operators
$c^\#_ic^\#_j$ with $i,j=1,2$. 
If we apply these generators to the basis vectors,
it is easy to see that Alice's observables respect the block decomposition
Eq.~(\ref{block-decomp-intro}) by acting trivially on Bob's subspace and
preserving the parity of Alice's subspace. For example,
$c^\#_ic^\#_j\Psi_{++}\in\2H_{++}$ for $i,j=1,2$.

Putting all these together, a general operator of Alice's local
observable algebra can be represented by the block matrix
\begin{equation}
A=\left(\begin{array}{cccc}
	A_+\otimes\I_+&&\\
	&A_+\otimes\I_-&\\
	&&A_-\otimes\I_+&\\
	&&&A_-\otimes\I_-\\
      \end{array}\right)
\end{equation}
with $2\times2$ matrices $A_\pm\in M_2(\7C)$.
By an analogous reasoning, a
general operator that belongs to Bob's local observable
algebra has the form
\begin{equation}
B=\left(\begin{array}{cccc}
	\I_+\otimes B_+&&\\
	&\I_+\otimes B_-&\\
	&&\I_-\otimes B_+&\\
	&&&\I_-\otimes B_-\\
      \end{array}\right)
\end{equation}
with $2\times2$ matrices $B_\pm\in M_2(\7C)$. This representation of the local
observable algebras shows directly the fact that Alice's operators commute with
Bob's.

We suggest here a notion for maximally entangled bipartite fermion states which
is related to the maximization of the entanglement of formation. In view of our
example above, we declare a state 
\begin{equation}
\label{max-ent-intro}
\varrho=\left(\begin{array}{cccc}
	p_{++}E_{++}&&&\\
	&p_{+-}E_{+-}&&\\
	&&p_{-+}E_{-+}&\\
	&&&p_{--}E_{--}\\
      \end{array}\right) \ ,
\end{equation}
where $E_{xy}$, $x,y=\pm$, are projections onto maximally entangled vectors
$\Omega_{xy}$, to be maximally entangled.

The example given above, is concerned with two fermion modes per site (two for
Alice and two for Bob) and the total dimension of the Hilbert space is $2^4$.
More generally, if each of Alice and Bob possesses $n$ fermion modes, then the
dimension is $2^{2n}$. Therefore the dimension of the Hilbert space of the
system increases exponentially with the number of fermion modes. It is
therefore highly desirable to find ways of treating at least some aspects of
large fermion systems without actually having to write out state vectors
component by component. Here the class of {\em quasifree fermion states} is a
suitable choice. They can be characterized by simple combinatorial data, which
do not grow exponentially but at most quadratic with the number of modes.
Quasifree states are determined by the two-point correlation functions, i.e.,
the expectation values $\expec{c_\4i^\#c_\4j^\#}$. For $2n$ fermion mode a
quasifree state is determined by a $4n \times 4n$ 
matrix which have to fulfill
suitable constraints that are related to the positivity of the density operator
for the state. Although the mathematical structure of quasifree states can be
treated easily, this class of states is sufficiently complex to tackle many
non-trivial physical systems, in particular to find ground states of interacting
spin chains with the help of the Jordan-Wigner transformation. 

To determine a quasifree
state, we arrange the expectation values $\expec{c_\4i^\#c_\4j^\#}$, $\4i,\4j=1, \ldots 2n$,
into the $4n\times4n$ 
hermitian matrix
\begin{equation}
S=\left(\begin{array}{cc}
        S_\4{AA}&S_\4{AB}\\
	S_\4{BA}&S_\4{BB}
       \end{array}\right) \ ,
\end{equation}
where $S_\4{BA}=S_\4{AB}^\adj$.
Any of the $2n\times2n$ 
submatrices $S_\4{XY}$, with
$\4X,\4Y=\4A,\4B$, has the structure
\begin{equation}
S_\4{XY}
= \left(\begin{array}{cc}
S_\4{XY}^{+-} & S_\4{XY}^{++} \\
S_\4{XY}^{--} & S_\4{XY}^{-+}
\end{array}\right)
\end{equation}
and the $n\times{}n$ 
submatrices $S_\4{XY}^{xy}$, with $x,y=+,-$,
are given by $(S_\4{AA}^{+-})_{jk}=\langle{c_j^\dagger{}c_k}\rangle$,
$(S_\4{AB}^{+-})_{jk}=\langle{c_j^\dagger{}c_{n+k}}\rangle$, and so on.
Here, $\expec{c_i^\#c_j^\#}:=\tr(\varrho_Sc_i^\#c_j^\#)$.
For the sake of clarity, for the case $n=2$, we fully list the off-diagonal block $S_\4{AB}$, which
plays {\em the} essential role in view of entanglement:
\begin{equation}
S_{\4A\4B}=\left(\begin{array}{cccc}
\expec{c_1^\adj c_3}&\expec{c_1^\adj c_4}&\expec{c_1^\adj c_3^\adj}&\expec{c_1^\adj c_4^\adj}\\
\expec{c_2^\adj c_3}&\expec{c_2^\adj c_4}&\expec{c_2^\adj c_3^\adj}&\expec{c_2^\adj c_4^\adj}\\
\expec{c_1 c_3}&\expec{c_1c_4}&\expec{c_1 c_3^\adj}&\expec{c_1 c_4^\adj}\\
\expec{c_2 c_3}&\expec{c_2c_4}&\expec{c_2 c_3^\adj}&\expec{c_2 c_4^\adj}
\end{array}\right) \; .
\end{equation}
The matrix $S$ is called the {\em covariance matrix} of the quasifree state and
we denote by $\varrho_S$ the corresponding density matrix. The diagonal blocks
of $S$ are related to the Alice-Alice and Bob-Bob correlations. The ordering of
the entries of $S$ is chosen in agreement with Araki's selfdual formalism, which
is explained in the next section.

The main result of this paper gives a complete characterization of maximally
entangled quasifree states for the case that Alice and Bob possess an
arbitrary number $n$ of fermion modes by giving an explicit form of the
covariance matrix. For two modes per site, the standard example for a maximally
entangled fermion state is given by the covariance operator
\begin{equation}
P=
\frac{1}{2}\left(\begin{array}{cc}
\openone_4 & \8i\openone_4 \\
-\8i\openone_4 & \openone_4
\end{array}\right) \; ,
\end{equation}
where here and in the following the symbol $\openone_n$ denotes the $n\times{}n$
identity matrix.
It is not difficult to check that $P$ is a projection which implies that the
quasifree state $\varrho_P$ is pure {\em on the full fermion algebra} that is
generated by $c_1^\#,c_2^\#,c_3^\#,c_4^\#$. The first thing one observes here
is that the diagonal blocks $P_\4{AA}=P_\4{BB}=\I/2$ are given by one-half
times the unit operator. Thus the state restricted to Alice's fermion operators
has covariance $\I/2$. As we shall prove, the quasifree state with covariance
$\I/2$ is the totally mixed state, i.e., the corresponding density matrix is
$\varrho_{\I/2}=\I/4$. Recall that Alice's fermion operators $c_1^\#,c_2^\#$
generate the algebra $M_4(\7C)$ of all complex $4\times4$ matrices. The fact
that the reduced density matrix is maximally mixed is the essential ingredient
to prove that the reduced density matrix $of \varrho_P$ to the subalgebra that
is generated by Alice's and Bob's observables has indeed the form of
Eq.~(\ref{max-ent-intro}).

\section{Bipartite fermion systems and the CAR algebra}
\label{sec-1}

Let us now consider the most general case, where Alice controls a finite set $\4A$ of $n$ fermion modes, whereas Bob possesses another set $\4B$ of $m$ fermion modes.
We denote by $\6F_\4A$ the full local algebra of Alice,
which consists of all possible linear combinations of products of
creation and annihilation operators $c_\4a^\adj,c_\4a$ belonging to her set of modes $\4a\in\4A$.
Alice's local observable algebra is instead given by the even part of $\6F_\4A$, which is denoted by $\6F_\4A^+$ and is given by linear combinations of products of an even number of Fermi operators.
The same notation is used for Bob, by an obvious replacement of the subscripts.
The full global algebra generated by Alice's and Bob's fermion operators together
is denoted by $\6F_\4{AB}$. The global observable algebra is the even part of $\6F_\4{AB}$ and is denoted by $\6F_\4{AB}^+$.
Finally, the even-even part of the global algebra, corresponding to the tensor product $\6F_\4A^+\otimes\6F_\4B^+$, is denoted by $\6F_\4{AB}^{++}$.
Similarly to the even part, one can also define the odd part of an algebra, e.g., $\6F_\4{A}^-,\6F_\4{B}^-,\6F_\4{AB}^-$. Note however that, differently from the even part, the odd part is only a linear subspace and not an algebra, as products of odd operators yield an even operator.

The bipartite structure is given by the natural separation of fermionic modes
relative to the (full) subalgebras $\6F_\4A$ and $\6F_\4B$.
Fermionic entanglement is based on the properties of correlation functions $\langle A B\rangle=\tr(\varrho AB)$, i.e., the expectation values of operator products $AB$ of a state $\varrho$, where $A$ belongs to Alice's fermion algebra $\6F_\4A$ and $B$ is an operator in Bob's fermion algebra $\6F_\4B$. In comparison to the standard entanglement theory, one is faced with the problem that the operators $A$ and $B$ do not necessarily commute. Namely, they anticommute whenever both $A$ and $B$ are odd operators, i.e., $A\in\6F_\4{A}^-$ and $B\in\6F_\4{B}^-$.
This can be seen as a non-local behaviour which makes the fermionic bipartite structure different from the standard tensor product. The appropriate description of this feature is provided by the twisted tensor product discussed in Sec. \ref{sec-twisted}.
On the other hand, in the physical situations where local parity is preserved, it is sufficient to consider the restriction to the even-even part of the algebra, where operators belonging to different subsystems commute and the standard tensor product can be used. This will be our assumption in Secs. \ref{sec-2} and \ref{sec-3}.
In the rest of this section we will introduce a convenient notation to deal with fermionic fields and fermionic Gaussian states.

\subsection{Araki's selfdual formalism}
\label{subsec-Araki}

In order to give a clear presentation, we make use of Araki's selfdual formalism which enables to write formulas in a more compact manner. The basic idea is to combine the creation and annihilation operators into one field operator by the linear combinations 
\begin{equation}
\B(f):=\sum_{\4j\in \4{AB}} f^+_\4j \ c_\4j+f^-_\4j  c^\adj _\4j \; .
\end{equation}
The $2(n+m)$ complex coefficients $f^\pm_\4j$ are put into one vector $f$ of a
Hilbert space $K_\4{AB}=\7C^n\oplus\7C^n\oplus \7C^m\oplus\7C^m$ which we call
the Hilbert space of the global one-particle system. Each vector $f\in K_\4{AB}$
can be decomposed into four blocks $f=(f^+_\4A,f^-_\4A,f^+_\4B,f^-_\4B)$ where
$f_\4A^\pm\in\7C^n$ are the one-particle components for Alice and
$f_\4B^\pm\in\7C^m$ are the one-particle components for Bob. Recall that Alice
possesses $n=|\4A|$ modes and Bob has $m=|\4B|$ modes under control. We
introduce a complex conjugation $\Gamma$ which allows to write the canonical
anticommutation relation in a more elegant form. Making use of our block
decomposition, the complex conjugation is an antiunitary operator that can be
represented by a $4\times4$ block matrix  
\begin{equation}
\label{complex-conjugation}
\Gamma
=\left(\begin{array}{cccc}
0&J_n&0&0\\
J_n&0&0&0\\
0&0&0&J_m\\
0&0&J_m&0
\end{array}\right) \; .
\end{equation}
where $J_k$ (in our case we have $k=n,m$) is the complex conjugation on $\7C^k$ that replaces each component within a vector by its complex conjugate. By using the anticommutation relations for the standard creation and annihilation operators $c^\adj _\4j,c_\4j$ allows to write the canonical anticommutation relation in the following compact form:
\begin{equation}
\label{selfdual-r1}
\{\B(f),\B(g)\}=\sprod{\Gamma f}{g}\I \; ,
\end{equation}
where $\sprod{\,}{\,}$ denotes the usual complex scalar products between vectors.
Moreover, the adjoint of $B(f)$ can simply be computed according to
\begin{equation}
\label{selfdual-r2}
\B(f)^\adj =\B(\Gamma f)\; .
\end{equation}
Indeed, the global fermion algebra $\6F_\4{AB}$ is completely determined by requiring that the field operator $\B(f)$ is complex linear in the one-particle vector $f$ and by the relations Eqs.~(\ref{selfdual-r1}) (\ref{selfdual-r2}). 


\subsection{Quasifree states}
\label{sec-3-2}
Since $\6F_\4{AB}$ is a finite dimensional full matrix algebra, the states on the bipartite fermion system are given by density operators $\varrho$ assigning to an operator $X$ the expectation value $\langle X\rangle=\tr(\varrho X)$.

A quasifree state $\varrho$, also called Gaussian fermion state, is completely
determined by the two-point correlation functions, i.e.,
the expectation values of 
$\tr(\varrho c_\4jc_\4i),\tr(\varrho c_\4j^\adj
c_\4i^\adj ),\tr(\varrho c_\4jc_\4i^\adj ),\tr(\varrho c_\4j^\adj c_\4i)$ 
Araki's selfdual approach
is very handy by putting all this expectation values into a single big
covariance matrix. Each quasifree state $\varrho=\varrho_S$ is in one-to-one
correspondence with a covariance matrix which is a linear operator
$S$ on $K_\4{AB}$ that is between $0$ and $\I$ ($0<S\leq \I$) with respect to
the operator ordering and which fulfils the additional constraint $S+\Gamma
S\Gamma=\I$. The expectation values of the state $\omega$ are related to its
covariance matrix $S$ by the following condition on the two-point correlation
function:
\begin{equation}\label{eq:twopoint}
\tr(\varrho_S \B(f)\B(g))=\sprod{\Gamma f}{Sg}\; .
\end{equation}
The entries of the matrix $S$ are related to pair correlation functions as described in Sec. \ref{sec-1}.
All higher correlation functions can be expressed in terms of sums of products
of two-point functions according to Wick theorem, where only the expectation values of an even product of Fermi field operators are non vanishing (i.e., $\varrho_S\in\6F_{\4A\4B}^+$). It is well known that a quasifree
state is pure if and only if its covariance matrix $S$ is a projection, called
{\em basis projection}. One simple example for a basis projection can be given
in terms of the $4\times4$ block matrix
\begin{equation}
\label{prod-state}
E:=\left(\begin{array}{cccc}
\I_n&0&0&0\\
0&0&0&0\\
0&0&\I_m&0\\
0&0&0&0
\end{array}\right)
\end{equation}
according to the four-block-decomposition of $K_\4{AB}$. Indeed, a
straightforward computation shows that  the condition $E+\Gamma E\Gamma=\I$ is
fulfilled. The expectation values for the standard creation and annihilation
operators are  $\tr(\varrho_E c_a^\adj c_a)=\tr(\varrho_E c_b^\adj
c_b)=1$ for $a\in\4A$ and $b\in\4B$. The expectation values of all other
possible combinations of Fermi operators are vanishing. In particular,
$\tr(\varrho_E\B(Q_\4A f)\B(Q_\4B h))=0$ which implies that $\varrho_E$ is a
product state, i.e., if we consider a product $AB$ of an operator $A$ from
Alice's local fermion algebra and an operator $B$ from Bob's local fermion
algebra, then the expectation value $\tr(\varrho_E AB)=\tr(\varrho_E
A)\tr(\varrho_E B)$ factorizes.

\section{Maximally entangled states on bipartite fermion systems}
\label{sec-2}

In the following we assume that Alice and Bob control the same number of
fermion modes $n=m$. Due to the bipartite nature of the fermion algebra
$\6F_\4{AB}$ there is a natural notion of product states. A state $\varrho$ on
$\6F_\4{AB}$ is called a product state if the expectation value $\tr(\varrho
AB)=\tr(\varrho_\4A A)\tr(\varrho_\4B B)$ factorizes for
any pair of operators $A,B$ belonging to Alice's and Bob's local fermion
algebras, respectively.
\footnote{Possible variations of this definition of product state are
discussed in Ref.~\cite{BanCirWol07}.}
The quasifree state given by Eq.~(\ref{prod-state}) is a particular example. As for
standard entanglement theory, a state on $\6F_\4{AB}$ is entangled if it can
not be written as a convex combination of product states
\cite{BanCirWol07,Mor04}. But what about maximally entangled states?
There is a natural point of view for discussing entanglement in fermion
systems. If we think of the observables that can be measured within Alice's and
Bob's local laboratories, we have to restrict to the even parts of the local
subalgebras, i.e., $\6F_\4A^+$ and $\6F_\4B^+$. In particular, each
correlation experiment for detecting the presence of entanglement can only be
built from observables that belong to the local even parts. As a consequence,
the {\em accessible} entanglement of a bipartite fermion state $\varrho$ is
the entanglement of the restriction to the even-even subalgebra
$\6F_{\4A\4B}^{++}$ which can be directly identified with the tensor
product $\6F_\4A^+\otimes\6F_\4B^+$.

As already mentioned, the algebra $\6F_\4{AB}$ can be identified with a full
matrix algebra of all linear operators on a Hilbert space with dimension
$2^{2n}$. We choose now an identification that is most compatible with the
bipartite nature, namely we identify $\6F_\4{AB}=\9B(\2H\otimes\2H)$ with the
matrix algebra of all linear operators on $\2H\otimes \2H$, where
$\2H:=F_-(\7C^n)$ is the antisymmetric Fock space over $\7C^n$. We identify the
set $\4A$ of labels for Alice's fermion modes with
$\4A=\{a_1,a_2,\cdots,a_n\}$
and the set of labels for Bob's fermion
modes with $\4B=\{b_1,\cdots,b_n\}$. Let $c^\adj _j, c_j$ denote the standard
creation and annihilation operators on $\2H$. Then Alice's fermion operators
act on $\2H\otimes \2H$ by $c_{a_j}=c_j\otimes\I$ and Bob's fermion operators
act on $\2H\otimes\2H$ by $c_{b_j}=\theta\otimes c_j$, where
$\theta=(-1)^{\boldsymbol N}$ is the parity of the particle number 
${\boldsymbol N}=\sum_{j=1}^n c_j^\adj c_j$ on $\2H$.
The Fock space $\2H=\2H_+\oplus\2H_-$ decomposes into the even and odd particle
number subspaces $\2H_\pm$ that correspond to the eigenspaces of $\theta$ with
eigenvalues $\pm 1$.
The projections onto $\2H_\pm$ are $\I_\pm=(\I\pm\theta)/2$ and clearly $\I=\I_+\oplus\I_-$.
Thus the tensor product $\2H\otimes \2H$ can be
expanded into a direct sum of four subspaces
$\2H\otimes\2H=(\2H_+\otimes\2H_+)\oplus(\2H_+\otimes\2H_-)\oplus(\2H_-\otimes\2H_+)\oplus(\2H_-\otimes\2H_-)$.
The local parity number operators for Alice and Bob are given by
$\Theta_\4A=\theta\otimes\I$ and $\Theta_\4B=\I\otimes\theta$. The even-even
subalgebra $\6F_\4{AB}^{++}$ consits of all operators $X$ that preserve the
local parity, i.e., $X$ commutes with $\Theta_\4A$ and $\Theta_\4B$. Therefore
the operator $X\in\6F_\4{AB}^{++}$ is block diagonal $X=X_{++}\oplus
X_{+-}\oplus X_{-+}\oplus X_{--}$. This yields the desired decomposition 
\begin{equation}
\label{decomp}
\6F_\4{AB}^{++}=\bigoplus_{a,b=\pm} \9B(\2H_a)\otimes\9B(\2H_b) ; .
\end{equation}
where $\2H_+$ and $\2H_-$ have the same dimension $2^{n-1}$.

If the global fermion system is prepared in a state $\varrho$, then the restriction $\varrho^{++}$ to the even-even part has a unique convex decomposition
\begin{equation}
\label{conv-decomp}
\varrho^{++}=\sum_{a,b=\pm} p_{ab} \ \sigma_{ab}
\end{equation}
where $\sigma_{ab}$ is a state on $\9B(\2H_a)\otimes\9B(\2H_b)$. We suggest here to characterize maximally entangled fermion states by the following definition:

\begin{defi}
\label{max-ent}
{\em
A bipartite fermion state $\varrho$ on $\6F_\4{AB}$ is called {\em maximally
entangled} if each state $\sigma_{ab}$ within the decomposition
Eq.~(\ref{conv-decomp}) is maximally entangled in the usual sense.}
\end{defi}

This point of view is related to searching for those states for which the
restriction $\varrho^{++}$ to the even-even part maximizes the entanglement of
formation $E_F$. Recall that it can be evaluated according to 
\begin{equation}
E_F(\varrho^{++})
=\inf_{\{q_x,\sigma_x|\varrho^{++}=\sum q_x\sigma_x\}}\sum_x q_x E(\sigma_x) \ ,
\end{equation}
where the infimum is taken over all possible decompositions $\varrho^{++}=\sum
q_x\sigma_x$ into pure states $\sigma_x$ and $E$ denotes the entropy of
entanglement.
If the state is maximally entangled according to Def.~\ref{max-ent},
the states $\sigma_{ab}$ inside decomposition (\ref{conv-decomp}), being
maximally entangled in the usual sense, must be pure.
Then, the direct sum structure of $\6F_\4{AB}^{++}$ implies
that the decomposition (\ref{conv-decomp}) is the only decomposition into pure
states and hence
\begin{equation}
E_F(\varrho^{++})
=\sum_{a,b=\pm} p_{ab} \ E(\sigma_{ab})=\log_22^{n-1}=n-1 \ .
\end{equation}
The density matrix $\sigma_{ab}$ is the projection onto a maximally
entangled vector $\Omega_{ab}$ in $\2H_a\otimes\2H_b$.
This naturally induces
a vector $\Omega=\oplus_{a,b=\pm}\sqrt{p_{ab}}\Omega_{ab}$ in
the full Hilbert space $\2H\otimes\2H$ which corresponds to a pure state
$\varrho=|\Omega\rangle\langle\Omega|$ on the global fermion algebra
$\9B(\2H\otimes\2H)=\6F_\4{AB}$.
Restricting the analysis to the accessible entanglement mentioned above,
yields a definition that only constrains the form of $\varrho^{++}$.
The global fermion state $\varrho$ is then not uniquely determined. It
is however natural to require it to be pure, as for usual maximally entangled
states. Therefore, we will assume each maximally entangled bipartite fermion
state to be pure on the global fermion algebra.

\subsection{The structure of maximally entangled quasifree states}
Concerning quasifree states, we are faced with the following problem: Is it possible to characterize explicitly the set of those covariance matrices which correspond to a maximally entangled fermion state? 

\begin{thm}
\label{thm-max-ent}
A quasifree pure state $\varrho_P$ on the global fermion algebra $\6F_\4{AB}$ with
covariance matrix $P$ is maximally entangled if and only if there exists a
unitary operator $U_\4{AB}$ that maps Bob's one-particle space $K_\4B$ onto
Alice's one-particle space $K_\4A$ and which anticommutes with the complex
conjugation $\Gamma_\4A U_\4{AB}=- U_\4{AB}\Gamma_\4B$ such that the covariance
operator is given by the $2\times2$ block matrix
\begin{equation}
\label{basis-projection}
P=
\frac{1}{2}
	\left(\begin{array}{cc}
	\I_\4A&U_\4{AB}\\
	U_\4{BA}&\I_\4B
	\end{array}\right)\; .
\end{equation}
\end{thm}
\begin{proof}
Suppose a maximally entangled quasifree fermion state $\varrho_P$ with covariance
matrix $P$ is given. Since $\varrho_P$ is pure, the covariance $P$ is a basis
projection and the corresponding density operator $\varrho_P$ is the projection
onto a vector $\Omega_P=\bigoplus_{a,b=\pm} \sqrt{p_{ab}}\Omega_{ab}$ with
maximally entangled vectors $\Omega_{ab}$. As a consequence, the restriction to
Alice's even subalgebra is a trace, i.e., for each $A_1,A_2\in\6F_\4{A}^+$ we
have $\tr(\varrho_P A_1A_2)=\tr(\varrho_P A_2A_1)$. Moreover, the restriction of
the quasifree state to Alice local fermion algebra $\6F_\4A$ is again a quasifree
state with covariance operator $Q_\4A P Q_\4A$, where $Q_\4A$ is the projection
onto Alice's one-particle space. Recall that each quasi free state is determined
by the two-point correlation function. According to Theorem~\ref{thm-app-1},
which we prove in the appendix, the covariance operator for a quasifree state
which is a trace on the even part must be $\I/2$. Thus we conclude that $Q_\4A P
Q_\4A=Q_\4A/2$. The same also holds for the restriction to Bob's even subalgebra
and we also find $Q_\4B P Q_\4B=Q_\4B/2$. With respect to the Alice-Bob split
$K_\4{AB}\cong K_\4A\oplus K_\4B$ into the local one-particle spaces, the
covariance operator can be written as a $2\times2$ block matrix
\begin{equation}
P=\frac{1}{2}\left(\begin{array}{cc}
\I_\4A&U_\4{AB}\\
U_\4{BA}&\I_\4B
\end{array}\right)\; .
\end{equation}
Here $U_\4{AB}$ is an operator from Bob's one-particle space $K_\4B$ to Alice's
one-particle space $K_\4A$ and $U_\4{BA}$ maps conversely $K_\4B$ to $K_\4A$.
Since $P$ is selfadjoint, the off-diagonal blocks
must fulfill $U_\4{AB}=U_\4{BA}^\adj $. Moreover, $P$ is idempotent, i.e.,
$P^2=P$ which implies
\begin{equation}
\begin{split}
P^2
	&=\frac{1}{4}
		\left(
			\begin{array}{cc}
			\I_\4A+U_\4{AB}U_\4{BA}&2U_\4{AB}\\
			2U_\4{BA}&\I_\4B+U_\4{BA}U_\4{AB}
			\end{array}
		\right)
\\
	&=\frac{1}{2}
		\left(
			\begin{array}{cc}
			\I_\4A&U_\4{AB}\\
			U_\4{BA}&\I_\4B
			\end{array}
		\right) \; .
\end{split}
\end{equation}
Comparing the diagonal blocks yields the two identities $U_\4{AB}U_\4{BA}=U_\4{AB}U_\4{AB}^\adj =\I_\4A$ and $U_\4{BA}U_\4{AB}=U_\4{AB}^\adj U_\4{AB}=\I_\4B$ which shows that $U_\4{AB}$ is a unitary operator. Finally we have to take care of the constraint $P+\Gamma P\Gamma=\I$ for $P$ being a valid covariance matrix. With respect to the Alice-Bob split of the one-particle space, the complex conjugation $\Gamma$ is block diagonal 
\begin{equation}
\Gamma
=
\left(\begin{array}{cc}
	\Gamma_\4A&0\\
	0&\Gamma_\4B
\end{array}\right)\; .
\end{equation}
Recall that by comparing this $2\times2$ block representation with the
$4\times4$ block representation Eq.~(\ref{complex-conjugation}) the complex
conjugation on Alice's (Bob's) one-particle space is given by the $2\times2$ block matrix
\begin{equation}
\Gamma_\4A
=
\left(\begin{array}{cc}
	0&J_n\\
	J_n&0
\end{array}\right)
\end{equation}
where $J_n$ is the natural complex conjugation on $\7C^n$. Expressed in terms of the Alice-Bob split,  the condition $P+\Gamma P\Gamma=\I$ becomes
\begin{equation}
\begin{split}
\Gamma P\Gamma
	&=\frac{1}{2}
		\left(
			\begin{array}{cc}
			\I_\4A&\Gamma_\4A U_\4{AB}\Gamma_\4B\\
			\Gamma_\4B U_\4{AB}^\adj \Gamma_\4A&\I_\4B
			\end{array}
		\right)
\\
	&=
	\frac{1}{2}
		\left(
			\begin{array}{cc}
			\I_\4A&-U_\4{AB}\\
			-U_\4{AB}^\adj &\I_\4B
			\end{array}
		\right)
	=\I-P \; .
\end{split}
\end{equation}
By comparing the off-diagonal blocks, we obtain the desired relation
$U_\4{AB}\Gamma_\4B=-\Gamma_\4A U_\4{AB}$. 

It remains to prove the converse. Suppose a unitary $U_\4{AB}$ is given such
that $U_\4{AB}\Gamma_\4A=-\Gamma_\4B U_\4{AB}$ holds. Then
Eq.~(\ref{basis-projection}) defines a basis projection $P$ that corresponds to a
pure state $\varrho_P=|\Omega_P\rangle\langle\Omega_P|$ with
$\Omega_P\in\2H\otimes\2H$. With respect to the direct sum decomposition into
even and odd particle number subspaces $\2H_\pm$ the vector $\Omega_P$ can be
written as $\Omega_P=\bigoplus_{ab}\sqrt{p_{ab}}\Omega_{ab}$ where $\Omega_{ab}$
are normalized vectors in $\2H_a\otimes\2H_b$.
Since a quasifree state vanishes on the odd part of the global fermion
algebra, the vector $\Omega_P$ has to be an eigenvector of $\Theta$. Thus either
$p_{+-}=p_{-+}=0$ or $p_{++}=p_{--}=0$.
Furthermore, the restriction of $\varrho_P$ to Alice's local
fermion algebra  $\6F_\4A$ is a trace
since, by construction,
$Q_\4APQ_\4A=Q_\4A/2$.
Since
$\6F_\4A\cong[\9B(\2H_+)\otimes\I]\oplus[\9B(\2H_+)\otimes\I]$, the restriction of
$\varrho_P$ to $\9B(\2H_\pm)\otimes\I$ is a trace. If $p_{+-}=p_{-+}=0$, then
the restriction of of $\varrho_P$ to 
\begin{equation}
\9B(\2H_\pm)\otimes\I\cong\left(\begin{array}{cc}
\9B(\2H_\pm)\otimes\I_+&0\\
0&\9B(\2H_\pm)\otimes\I_-
\end{array}\right)
\end{equation}
is supported on the blocks $\9B(\2H_+)\otimes\I_+$ and $\9B(\2H_-)\otimes\I_-$.
On the other hand, if $p_{++}=p_{--}=0$, then the restriction of of $\varrho_P$
is supported on the blocks $\9B(\2H_+)\otimes\I_-$ and $\9B(\2H_-)\otimes\I_+$.
Therefore, the restriction of the pure state
$\sigma_{ab}=|\Omega_{ab}\rangle\langle\Omega_{ab}|$ is a trace on
$\9B(\2H_a)\otimes\I_b$ which implies that $\Omega_{ab}$ is a maximally
entangled vector in $\2H_a\otimes\2H_b$ and $\varrho_P$ is a maximally entangled
fermion state in the sense of Definition~\ref{max-ent}.
\end{proof}

In summary, the theorem tells us that all maximally entangled fermion states
can be characterized by a unitary operator $U_\4{AB}$ from Bob's one-particle
space onto Alice's one-particle space that anticommute with the complex
conjugation $\Gamma_\4A U_\4{AB}=-U_\4{AB}\Gamma_\4B$. Since, according to our
assumptions, Alice's and Bob's one-particle spaces have the same dimension $2n$
they can be canonically identified $K_\4A= K_\4B=\7C^{2n}$. The simplest choice
for the unitary $U_\4{AB}$ is just given by $U_\4{AB}=\8i\I_{2n}$, where
$\I_{2n}$ is the unit operator on $\7C^{2n}$. The global one particle space is
given by $\7C^{2n}\oplus\7C^{2n}$ and the basis projection simply looks like 
\begin{equation}
\label{standard-max}
P_\8{st}:=\frac{1}{2}
		\left(
			\begin{array}{cc}
			\I_{2n}&\8i\I_{2n}\\
			-\8i\I_{2n}&\I_{2n}
			\end{array}
		\right) \; .
\end{equation}
We call the corresponding quasifree fermion state the {\em standard maximally entangled fermion state}.

\subsection{Local Bogolubov transformations}
As we will see soon, each quasifree maximally entangled fermion state can be
transformed into the standard maximally entangled fermion state by {\em local
Bogolubov transformations}. A Bogolubov transformation is given by a unitary
operator $V$ on $K_\4{AB}$ that commutes with the complex conjugation $\Gamma
V=V\Gamma$. Indeed, if we transform the Fermi field operators by
$\B(f)\mapsto\B(Vf)$ the anticommutation relations are preserved. Namely, we
have $\{\B(Vf),\B(Vh)\}=\sprod{\Gamma Vf}{Vh}\I=\sprod{V\Gamma
f}{Vh}\I=\sprod{\Gamma f}{h}\I$. This implies that $\B(f)\mapsto\B(Vf)$ extends
to a so called {\em Bogolubov automorphism} $\beta_V$ of the fermion algebra
$\6F_\4{AB}$. Since we are concerned with finite dimensions, there exists a
unitary operator $\U_V$ such that $\beta_V(\B(f))=\B(Vf)=\U_V^\adj\B(f)\U_V$.
Bogolubov automorphisms map quasifree states onto quasifree state. Indeed, a
quasifree state $\varrho_S$ with covariance $S$ is mapped to the quasifree state
$\U_V^\adj\varrho_S \U_V=\varrho_{VSV^\adj}$ with covariance $VSV^\adj$
\cite{Ara70,Ara87}.

A {\em local} Bogolubov transformation is block-diagonal with respect to the Alice-Bob split of the one-particle space, i.e., there are unitary operators $V_\4A$ on $K_\4A$ and $V_\4B$ on $K_\4B$ such that 
\begin{equation}
V=\left(
			\begin{array}{cc} 
			V_\4A&0\\
			0&V_\4B
			\end{array}
		\right)
\end{equation}
where $V_\4A$ and $V_\4B$ commute with the local complex conjugations
$\Gamma_\4A$ and $\Gamma_\4B$, i.e., they are Bogolubov transformations on the
local one-particle spaces $K_\4A$ and $K_\4B$, respectively.

By construction, a local Bogolubov transformation preserves Alice's and Bob's
even subalgebras. Namely, the block-diagonal structure of $V$ implies that an
operator $\B(f_\4A)$ of Alice's local fermion algebra is mapped to
$\B(Vf_\4A)=\B(V_\4A f_\4A)$ which is also contained in Alice's local fermion
algebra. The corresponding statement also holds for Bob.
Thus we can restrict the Bogolubov automorphism $\beta_V$ to an automorphism
$\beta_V^{++}$ to the even-even subalgebra
$\6F^{++}_\4{AB}$ which, as mentioned above, can be directly identified with the
tensor product $\6F_\4A^+\otimes\6F_\4B^+$ of Alice's even part and Bob's even
part. Then, the automorphism $\beta_V^{++}$ splits into a tensor product
$\beta_V^{++}=\beta_{V_\4A}^+\otimes\beta_{V_\4B}^+$,
where $\beta_{V_\4A}^+$ and $\beta_{V_\4B}^+$ are the restrictions of the
Bogolubov automorphisms $\beta_{V_\4A}$ and $\beta_{V_\4B}$ to Alice's and Bob's
even subalgebra. Hence a local Bogolubov transformation corresponds to a local
operation in the usual sense. 

If we apply a local Bogolubov transformation $V$ to the standard maximally
entangled state whose basis projection is given by the $P_\8{st}$ of
Eq.~(\ref{standard-max}), we obtain a new maximally entangled state with basis
projection $VP_\8{st}V^\adj$ which has the $2\times2$ block decomposition
\begin{equation}
VP_\8{st}V^\adj:=\frac{1}{2}
		\left(
			\begin{array}{cc}
			\I_{2n}&\8iV_\4A V_\4B^\adj\\
			-\8i V_\4B V_\4A^\adj&\I_{2n}
			\end{array}
		\right) \; .
\end{equation}

Let $U_\4{AB}$ be a unitary that anticommutes with the complex conjugation and therefore determines a maximally entangled fermion state. Then $-\8iU_\4{AB}$ commutes with the complex conjugation and 
\begin{equation}
W=\left(
			\begin{array}{cc}
			-\8iU_\4{AB}&0\\
			0&\I_{2n}
			\end{array}
		\right)
\end{equation}
is a local Bogolubov transformation that acts on Bob's system alone. If we apply it to the standard maximally entangled state, then we find 
\begin{equation}
\label{standard-max-evol}
WP_\8{st}W^\adj:=\frac{1}{2}
		\left(
			\begin{array}{cc}
			\I_{2n}&U_\4{AB}\\
			U_\4{AB}^\adj&\I_{2n}
			\end{array}
		\right) 
\end{equation}
which proves the following statement:

\begin{thm}
Up to local Bogolubov transformations, there exists a unique maximally entangled quasifree fermion state.
\end{thm}

\subsection{Generating maximally entangled states}
Let us now consider the following one particle Hamilton operator $H$ on $K_\4{AB}$
\begin{equation}
H:=\left(\begin{array}{cccc}
0&0&0&\I_n\\
0&0&-\I_n&0\\
0&-\I_n&0&0\\
\I_n&0&0&0
\end{array}\right) \; .
\end{equation}
$K_\4{AB}$
Here $K_\4{AB}$ is given by the decomposition 
$K_\4{AB}\cong\7C^n\oplus\7C^n\oplus\7C^n\oplus\7C^n$ where the first two
summands are the one-particle space of Alice and the last two blocks belong to
Bob. 
It is clear that $H$ is a reflection that anticommutes with the complex conjugation $\Gamma$, i.e., we have $H^2=\I$ and $H\Gamma=-\Gamma H$. Thus for each $t\in\7R$ the unitary operator $\exp(\8itH)=\cos(t)\I+\8i\sin(t)H$ is a Bogolubov transformation. 

The one-particle Hamilton operator $H$ induces a second-quantized Hamilton operator $\boldsymbol H$ that acts on the antisymmetrized Fock space. The relation between the second-quantized  $\boldsymbol H$ and the one-particle operator $H$ is given by the commutator relation 
\begin{equation}
\label{eofm}
\left[{\boldsymbol H} ,\B(f)\right]=\B(Hf)
\end{equation}
which is the Heisenberg equation of motion. In order to express $\boldsymbol H$ in terms of Fermi field operators, we choose a orthonormal basis of $K_{AB}$ of $4n$ vectors $e^\pm_{\4Aj},e^\pm_{\4Bj}$ with $j=1,\cdots,n$ such that $\Gamma e^+_{\4Aj}=e^-_{\4Aj}$. The operator $H$ can be written in terms of this basis according to 
\begin{equation}
H=\sum_j\ket{e^+_{\4Aj}}\bra{e^-_{\4Bj}}-\ket{e^-_{\4Aj}}\bra{e^+_{\4Bj}}+ \mbox{h.c.} \; .
\end{equation}
The anticommutation relations Eq.~(\ref{selfdual-r1}) can be used to compute commutators of the following form: 
\begin{equation}
\begin{split}
[\B(e_1)^\adj\B(e_2),&\B(f)]
\\
&=\B([\Gamma\ket{e_1}\bra{e_2}\Gamma-\ket{e_2}\bra{e_1}]f) \; ,
\end{split}
\end{equation}
where $e_1,e_2,f$ are vectors in the one-particle space $K_\4{AB}$. This relation can directly be employed to check that the operator 
\begin{equation}
{\boldsymbol H}:=\sum_{j=1}^n \B(e^-_{\4Aj})^\adj\B(e^+_{\4Bj})+\B(e^+_{\4Bj})^\adj\B(e^-_{\4Aj}) 
\end{equation}
fulfills the equation of motion Eq.~(\ref{eofm}). Thus, we obtain a one-parameter group of Bogolubov automorphisms which is generated by $\boldsymbol H$ and which we regard now as the time evolution of the global fermion system. Recall, that the exponential form of Eq.~(\ref{eofm}) is given by 
\begin{equation}
\label{exp-eofm}
\exp(-\8it{\boldsymbol H})\B(f)\exp(\8it{\boldsymbol H})=\B\left(\exp(\8it H)f\right) \; .
\end{equation}

Now we apply this time evolution to the quasifree product state given by Eq.~(\ref{prod-state}) whose basis projection is given by $E_0:=E$. Then after time $t$ we have created a quasi free state that corresponds to the basis projection $E_t=\exp(-\8itH)E\exp(\8itH)$. By expanding the exponentials and by making use of the relation $HE_0H=\I-E_0$ we obtain 
\begin{equation}
\begin{split}
E_t=\cos(t)^2E_0+\sin(t)^2&(\I-E_0)
\\+&\8i\sin(t)\cos(t)[E_0,H] \; . 
\end{split}
\end{equation}
In terms of the $4\times4$ block decomposition, we find for the basis projection at time $t$:
\begin{equation}
E_t:=\left(\begin{array}{cccc}
	c_t^2 \I_n& 0& 0&-\8i s_t c_t \I_n\\
	0 &s_t^2 \I_n& \8i s_t c_t\I_n&0\\
	0&-\8i s_t c_t\I_n& c_t^2\I_n&0\\
	-\8i s_t c_t\I_n&0&0& s_t^2\I_n
\end{array}\right)
\end{equation}
where $c_t=\cos(t)$ and $s_t=\sin(t)$. Suppose Alice and Bob prepare their fermion  systems individually such that the global system is in the quasifree pure product state with basis projection $E_0$. Moreover, Alice's and Bob's fermions interact with each other by the dynamics $\beta_{\exp(\8itH)}$. After the interaction time $t=\pi/4$ the global fermion system is in the quasifree state that is given by the basis projection
\begin{equation}
E_{\pi/4}:=\frac{1}{2}\left(\begin{array}{cccc}
	\I_n& 0& 0&\8i\I_n\\
	0 &\I_n& \8i\I_n&0\\
	0&-\8i\I_n&\I_n&0\\
	-\8i\I_n&0&0&\I_n
\end{array}\right)
\end{equation} 
which is the basis projection of a maximally entangled fermion state.

\section{Spin chains and maximally entangled quasifree fermions: An instructive example}
\label{sec-3}
In this section we consider the example of fermionic quasi-free maximally
entangled state obtained from the Jordan-Wigner transformation of a maximally
entangled state in a spin chain. The spin state vector $\ket{\chi}$
defined below is related to the important class of isotropic states
\cite{Hor99,Rai99-01}, which can indeed be decomposed in terms of the identity and
the operator $\ket{\chi}\bra{\chi}$. We will show that once written in terms of
fermion operators, the considered state is a quasi-free state whose
covariance matrix exactly satisfies the conditions stated in Theorem
\ref{thm-max-ent}.

We consider two blocks of spins, each block containing $n$ contiguous spins. By
numbering the sites starting from the first spin of the left, the block
$\4A=\{1,2,\cdots, n\}$ belongs to Alice, whereas the block
$\4B=\{n+1,n+2,\cdots, 2n\}$ belongs to Bob. We construct the state vector
$\ket{\chi}$ as a tensor product of maximally entangled spin pairs, such that
the $(n+1-\nu)$-th spin of the left block is entangled with the $\nu$-th spin of
the right block, $\nu=1,\dots,n$. Fig.~\ref{fig:01} illustrates this situation
for $n=7$. Explicitly, the vector reads
\begin{equation}
\ket{\chi} = \frac{1}{\sqrt{2^n}}\prod_{\nu=1}^n
(1+\sigma_{n+1-\nu}^+\sigma_{n+r+\nu}^+)\ket{\downarrow\dots\downarrow} \ ,
\end{equation}
where $\sigma_j^+$ is the usual Pauli raising operator of the $j$-th site.
In the following we will drop the normalization factor $2^{-n/2}$.

The Jordan-Wigner transformation is usually defined as 
\begin{equation}
\sigma_j^+ =(-1)^{\sum_{k<j}n_k}c_j^\adj \, , 
\end{equation}
where $n_k=c_k^\dagger c_k$. By substituting in
the above expression for $\ket{\chi}$ one gets
\begin{equation}
\ket{\chi} = \prod_{\nu=1}^n
[1+c_{n+1-\nu}^\dagger s(\nu) c_{n+\nu}^\dagger]\ket{0} \ ,
\end{equation}
where $s(\nu)=\prod_{n+1-\nu<j<n+\nu}(-1)^{n_j}$ (note that
$c_j(-1)^{n_j}=c_j$).
It is now possible to get rid of the `string' terms $s(\nu)$ by bringing them to
the right and applying them to the vacuum $\ket{0}$, so that
$s(\nu)\ket{0}=\ket{0}$. Indeed $[s(\nu),c_j^\dagger]=0$ whenever the site $j$ is
not contained in the string $s(\nu)$ and this condition is always satisfied in
the above expression for $\ket{\chi}$. Hence all the string terms can be
eliminated from the state, which just reads
\begin{equation}
\ket{\chi} = \prod_{\nu=1}^n
(1+c_{n+1-\nu}^\dagger c_{n+\nu}^\dagger)\ket{0} =
\prod_{\nu=1}^n e^{c_{n+1-\nu}^\dagger c_{n+\nu}^\dagger}\ket{0} \ .
\end{equation}
The latter expression puts in evidence that $\ket{\chi}$ is a
coherent state (with respect to properly defined fermion operators), namely
\begin{equation}\label{eq:coherent}
\ket{\chi} = 
e^{\frac{1}{2}\sum_{jk}c^\dagger_j G_{jk}c^\dagger_k}\ket{0}
\ ,
\end{equation}
where $j,k=1,\dots,2n$ and $G=i\sigma_y\otimes\mathbb{H}_n$.
Here $\mathbb{H}_n$ is the
trivial Hankel matrix obtained by reversing the $n\times{n}$ identity matrix,
i.e., $(\mathbb{H}_n)_{jk}=\delta_{j+k,n+1}$.
Note that $G$ is antisymmetric and orthogonal, $G^T=-G=G^{-1}$.

The covariance matrix for a state of the form Eq.~(\ref{eq:coherent}),
where $G$ is any real antisymmetric matrix,
can be obtained from the correlations
$\langle{c_j^\dagger{}c_k^\dagger}\rangle=
(T-T^T)_{jk}/4=-\langle{c_jc_k}\rangle$ and
$\langle{c_j^\dagger{}c_k}\rangle=(2\cdot\openone-T-T^T)_{jk}/4$, where
$T=(\openone+G)/(\openone-G)$.
These expressions can be obtained by using the standard techniques employed to
diagonalize quadratic fermion Hamiltonians (see, e.g.,
Ref.~\cite{Pesch03}). In our case a simple calculation shows that $T=G$, so
that
\begin{eqnarray}
\langle{c_j^\dagger{}c_k^\dagger}\rangle & = & \frac{G_{jk}}{2} \ , \\
\langle{c_j^\dagger{}c_k}\rangle & = & \frac{\delta_{jk}}{2} \ .
\end{eqnarray}

We recall that in the operators
$\bm{B}(f)$ the vector $f$ is ordered as
$f=(f_\4A^+,f_\4A^-,f_\4B^+,f_\4B^-)$ and that the
expectation value of the product $\bm{B}(f)\bm{B}(g)$
is given by Eq.~(\ref{eq:twopoint}).
The $4n\times4n$ matrix $S$ is given by
\begin{equation}
S
= \left(\begin{array}{cc}
S_\4{AA} & S_\4{AB} \\
S_\4{BA} & S_\4{BB}
\end{array}\right) \ ,
\end{equation}
where the structure of the $2n\times2n$ submatrices $S_\4{XY}$, with
$\4X,\4Y=\4A,\4B$, has already been described in Sec.~\ref{overview}.
Therefore, for the Gaussian state $|\chi\rangle$ introduced above one has
\begin{equation}
S = \frac{1}{2}
\left(\begin{array}{cccc}
\openone_n & & & \mathbb{H}_n \\
 & \openone_n & -\mathbb{H}_n & \\
 & -\mathbb{H}_n &\openone_n &  \\
\mathbb{H}_n & & & \openone_n
\end{array}\right) \ .
\end{equation}
Defining $U_\4{AB}=2S_\4{AB}$, it is easy to check that the matrix $U_\4{AB}$
satisfies the condition $\Gamma_\4AU_\4{AB}=-U_\4{AB}\Gamma_\4B$.

\section{Bipartite structure and the untwisting}
\label{sec-twisted}

In this section we describe the twisted tensor product structure which is
necessary to study entanglement in the global fermion algebra.
The bipartite nature of the fermion system can be revisited in terms of the
projections $Q_\4A$ and $Q_\4B$ onto Alice's and Bob's one-particle spaces
$K_\4A:=Q_\4AK_\4{AB}$ and $K_\4B:=Q_\4AK_\4{AB}$, respectively. With respect to
the $4\times4$ block decomposition, Alice's projection is given by  
\begin{equation}
Q_\4A:=\left(\begin{array}{cccc}
\I_n&0&0&0\\
0&\I_n&0&0\\
0&0&0&0\\
0&0&0&0
\end{array}\right)
\end{equation}
and the projection onto Bob's one particle space is the complementary
projection $Q_\4B:=\I-Q_\4A$. Alice's local fermion algebra $\6F_\4A$ is
generated by the operators $\B(Q_\4Af)$ and, analogously, Bob's local fermion
algebra $\6F_\4B$ is generated by the operators $\B(Q_\4Bf)$.  By construction,
the projection $Q_\4A$ (and therefore $Q_\4B=\I-Q_\4A$) commutes with the
complex conjugation which implies with help of Eq.~(\ref{selfdual-r1}):
$\{\B(Q_\4Af),\B(Q_\4Bh)\}=\sprod{\Gamma Q_\4Af}{Q_\4Bh}\I=\sprod{\Gamma
f}{Q_\4AQ_\4Bh}\I=0$. Therefore, the Alice's and Bob's fermion operators
mutually anticommute and the product of Alice's fermion operator $\B(Q_\4Af)$
and Bob's fermion operator $\B(Q_\4Bh)$ is not a tensor product, as it is the
case for (finite dimensional) commuting matrices. 

As a matrix algebra, the global fermion algebra $\6F_\4{AB}$ can be identified with the tensor product $\6F_\4A\otimes\6F_\4B$. In the following, this isomorphism, which we denote here by $\alpha$, is called the {\em untwisting}. It acts on fermi operators $\B(f)\in\6F_\4{AB}$ according to 
\begin{equation}
\label{jordan-wigner}
\alpha(\B(f)):=\B(Q_\4Af)\otimes\I + \Theta_\4A\otimes \B(Q_\4Bf) \ ,
\end{equation}
where $\Theta_\4A\in\6F_\4A$ is the reflection (selfadjoint unitary) that is
uniquely
determined by $\Theta_\4A \B(f)=\B(R_\4Af)\Theta_\4A$ and
$R_\4A=-Q_\4A+Q_\4B=\I-2Q_\4A$ is the reflection that changes the sign on Alice's
one-fermion space $K_\4A$. We mention here that $\Theta_\4A$ corresponds to a
string of $\sigma_3$-Pauli operators, as in the case of the Jordan-Wigner
transformation. 
valuating the untwisting $\alpha$ on a
product of Alice's and Bob's fermion operators gives 
\begin{equation}
\alpha(\B(f_\4A)\B(f_\4B))=\B(f_\4A)\Theta_\4A\otimes\B(f_\4B)  \; .
\end{equation}
Due to the occurrence of the operator $\Theta_\4A$ on the right hand side, the operator product $\B(f_\4A)\B(f_\4B)$ on the left hand side is called a ``twisted tensor product" of $\B(f_\4A)$ and $\B(f_\4B)$ which is ``untwisted'' by $\alpha$ to a tensor product on the right hand side.

We note that the even part $\6F_\4A^+$ of Alice's fermion algebra is precisely the fixpoint algebra under the map $A\mapsto\theta_\4A(A):=\Theta_\4A A\Theta_\4A$, i.e., an operator $A$ belongs to $\6F_\4A^+$ if and only if $\theta_\4A(A)=A$. The analogous result is also true for Bob's fermion algebra. Obviously, the even part $\6F_\4{AB}^+$ of the global fermion algebra consists of all operators that are invariant under the map  $F\mapsto\Theta F\Theta=F$, with $\Theta:=\Theta_\4A\Theta_\4B$. 

The essential property of the untwisting $\alpha$ is related to the odd/even
nature of Alice/Bobs fermion operators. Namely, for each of Alice's fermion
operators $A\in\6F_\4A$, the identities 
\begin{equation}
\begin{split}
&\alpha(AB_+)=A\otimes B_+
\\
&\alpha(AB_-)=A\Theta_\4A\otimes B_- \; .
\end{split}
\end{equation}
are valid, where $B_\pm\in\6F_\4B^\pm$ is any even (odd) fermion operator on Bob's side.

The twisted tensor product is hence the suitable tool to extend the analysis of entanglement to the global fermion algebra. Such an extension would be clearly desirable for those systems where local parity conservation cannot be assumed, in particular for general fermionic systems obtained from the Jordan-Wigner transformation of spin chain Hamiltonians.

\section{Conclusions}

In this paper, we have introduced a definition of fermionic maximally entangled
states based on the restriction to the even-even part of the algebra. Such a
restriction is natural as far as one is interested in the {\em physical}
operators (observables) of the two considered subsystems, which implies
conservation of local parity. We have then analyzed the structure of fermionic
maximally entangled states which are Gaussian with respect to the {\em global}
fermion algebra. The preparation of the state is indeed assumed to be
constrained by global parity only, as for observables of the total system.
Following this route, we have obtained a necessary and sufficient condition on
the covariance matrix for the Gaussian state to be maximally entangled. We have
also showed such a state to be unique up to local Bogolubov transformations. On
the other hand, we have provided a unitary transformation which maps a Gaussian
product state into a Gaussian maximally entangled state. An explicit example
relevant to spin chains has also been discussed. Finally, we have showed how to
extend the analysis of entanglement beyond the restriction to the even-even part
by introducing the twisted tensor product of fermionic subspaces. The latter structure is useful to study general fermionic systems obtained from the Jordan-Wigner transformation of lattice spins. The investigation of this context will be the subject of future work.

\begin{appendix}
\section{Traces on fermion algebras}
\label{app}
We consider the fermion algebra $\6F$ that is generated by the fermion field operators $\B(f)$ with $f\in K\cong\7C^n\oplus\7C^n$. Recall that the fermion algebra can be identified with the algebra $\9B(\2H)=\6F$ of all linear operators on the antisymmetric Fock space $\2H=F_-(\7C^n)$. On $\9B(\2H)$ there exists a unique normalized trace which is given by the density operator $2^{-n}\I$. 

\begin{lem}
\label{lem-app-1}
The normalized trace on the fermion algebra $\6F$ is the quasifree state with covariance $S=\I/2$, i.e., the density operator is given by $\varrho_{\I/2}=2^{-n}\I$.
\end{lem}
\begin{proof}
Let $\varrho_{\I/2}$ be the quasifree state that is given by the covariance $\I/2$. To prove that $\varrho_{\I/2}$ is a trace, we have to show that the correlation functions of even products of fermion fields are invariant under the cyclic shift, i.e., we have to prove the identity
\begin{equation}
\label{eq-1}
\begin{split}
\tr(\varrho_{\I/2}&\B(f_1)\B(f_2)\cdots\B(f_{2n}))
\\
=&\tr(\varrho_{\I/2}\B(f_{2n})\B(f_1)\cdots\B(f_{2n-1}))
\end{split}
\end{equation}
Let $Q_n\subset S_{2n}$ be the set of all permutations $q$ on $\{1,\cdots,2n\}$ that fulfill for each $k=1,2,3,\cdots, n$ the constraints
\begin{equation}
\begin{split}
&q(2k-1)<q(2k+1)
\\
&q(2k-1)<q(2k) \; .
\end{split}
\end{equation}
For our purpose, we introduce the following more handy representation for permutations 
\begin{equation}
\begin{split}
q=\left(\begin{array}{ccccc} 
q(1)&q(3)&\cdots&q(2n-1)\\
q(2)&q(4)&\cdots&q(2n)
\end{array}\right)
\end{split}
\end{equation}
where the first row contains the odd and the second row contains the even positions. The constraint for a permutation to be a member of $Q_n$ can be reformulated in this representation: (1) The entries of the first row are increasing from left to right. (2) In each column the first entry is smaller than the second.  By Wick's theorem, the left-hand side of Eq.~(\ref{eq-1}) can be expressed in terms of the two-point correlation functions by 
\begin{equation}
\label{wick-theo}
\begin{split}
\tr(\varrho_{\I/2}&\B(f_1)\B(f_2)\cdots\B(f_{2n}))
\\
&=
2^{-n}\sum_{q\in Q_n}\epsilon_q \prod_{k=1}^n\sprod{\Gamma f_{q(2k-1)}}{f_{q(2k)}}
\end{split}
\end{equation}
where $\epsilon_q$ is the sign of the permutation $q$. Let $\tau=(2 \ 3 \ \cdots \ 2n \ 1)$ be the cyclic shift, then we obtain for the right hand side
\begin{equation}
\begin{split}
\tr(\varrho_{\I/2}&\B(f_{2n})\B(f_2)\cdots\B(f_{2n-1}))
\\
&=
2^{-n}\sum_{q\in Q_n}\epsilon_q \prod_{k=1}^n\sprod{\Gamma f_{\tau q(2k-1)}}{f_{\tau q(2k)}}
\end{split}
\end{equation}
The problem is now, that the permutation $\tau q$ is not contained in $Q_n$, namely the constraints are violated at the position $i:=q^{-1}(2n)$. If we compose the cyclic shift $\tau$ with the permutation $q$ we obtain for $\tau q$:
\begin{equation}
\begin{split}
\left(\begin{array}{ccccc}
q(1)+1&\cdots&q(i-1)+1&\cdots&q(2n-1)+1\\
q(2)+1&\cdots&1&\cdots&q(2n)+1
\end{array}\right)
\end{split}
\end{equation}
The cyclic shift increases the value of the entry by one except the position where the value is equal to $2n$ which must be an even position $i=q^{-1}(2n)$. Thus, the entries of the first row are still increasing from he left to the right, but at position $i$ the first entry is now {\em smaller} than the second. This can be cured by introducing the permutation $\varsigma_i:=(i \ i-1\ 1 \ 2 \ \cdots \ i-2 \ i+1 \ \cdots \ 2n)$. Then we build the permutation $\kappa q:=\tau q \varsigma_i$ which can be represented as 
\begin{equation}
\begin{split}
\left(\begin{array}{ccccc}
1&q(1)+1&\cdots&q(2n-1)+1\\
q(i-1)+1&q(2)+1&\cdots&q(2n)+1
\end{array}\right) \; .
\end{split}
\end{equation}
By exchanging the two entries in column $i/2$ and shuffling it to the first position we have succeed in fulfilling the conditions (1) and (2). In fact, we have constructed a bijective map $\kappa\mathpunct:q\mapsto \tau q \varsigma_{q^{-1}(2n)}$ that acts on the set of allowed permutations $Q_n$.

The product term for the permutation $\tau q$ within the expansion Eq.~(\ref{wick-theo}) can be expressed in terms of the permutation $\kappa q$ by 
\begin{equation}
\begin{split}
	\prod_{k=1}^n&\sprod{\Gamma f_{\tau q(2k-1)}}{f_{\tau q(2k)}} 
	\\
	&= 
	\sprod{\Gamma f_{q(i-1)+1}}{f_1} \prod_{k=2}^n\sprod{\Gamma f_{\kappa q(2k-1)}}{f_{\kappa q(2k)}}
	\\
	&=
	\prod_{k=1}^n\sprod{\Gamma f_{\kappa q(2k-1)}}{f_{\kappa q(2k)}} \; .
\end{split}
\end{equation}
where we have used the fact that the complex bilinear form $(f,h)\mapsto\sprod{\Gamma f}{h}=\sprod{\Gamma h}{f}$ is symmetric. This implies the desired identity
\begin{equation}
\begin{split}
\tr(\varrho_{\I/2}&\B(f_{2n})\B(f_2)\cdots\B(f_{2n-1}))
\\
	&=
	2^{-n}\sum_{q\in Q_n}\epsilon_q \prod_{k=1}^n\sprod{\Gamma f_{\kappa q(2k-1)}}{f_{\kappa q(2k)}}
	\\
	&=
	2^{-n}\sum_{q\in Q_n}\epsilon_{\kappa^{-1}q} \prod_{k=1}^n\sprod{\Gamma f_{q(2k-1)}}{f_{q(2k)}}
	\\
	&=
	2^{-n}\sum_{q\in Q_n}\epsilon_{q} \prod_{k=1}^n\sprod{\Gamma f_{q(2k-1)}}{f_{q(2k)}}
	\\
	&=
	\tr(\varrho_{\I/2}\B(f_1)\B(f_2)\cdots\B(f_{2n})) \; .
\end{split}
\end{equation}
Here we have used the fact that the map $\kappa$ preserves the sign of permutations $\epsilon_{\kappa q}=\epsilon_q$. The sign of the cyclic shift of a set with $2n$ elements is $\epsilon_\tau=-1$. Namely, let $\sigma_k$ be the transposition that exchanges $k$ and $k+1$ we observe that $\tau=\sigma_{2n-1}\sigma_{2n-2}\cdots\sigma_1$ is a product of $2n-1$ transpositions. Furthermore, the permutation  
$\varsigma_i=(\sigma_1\sigma_2\cdots\sigma_{i-1})(\sigma_1\cdots\sigma_{i-2})$ can be decomposed into a product of $2(i-2)+1$ transpositions which implies $\epsilon_{\varsigma_i}=-1$. This yields $\epsilon_{\kappa q}=\epsilon_\tau\epsilon_q\epsilon_{\varsigma_i}=\epsilon_q$. The uniqueness of the normalized trace for full finite dimensional matrix algebras implies that the density matrix of the quasifree state $\varrho_{\I/2}$ is 
\begin{equation}
	\varrho_{\I/2}=2^{-n}\I 
\end{equation}
which concludes the proof.
\end{proof}

For the proof of Theorem~\ref{thm-max-ent} we need to consider all possible traces on the even subalgebra $\6F^+$ which coincides with $\9B(\2H_+)\oplus \9B(\2H_-)$ where $\2H_\pm$ is the subspace corresponding to even/odd particle number. Since on each full matrix algebra there is -- up to normalization -- a unique trace, all normalized traces on $\6F^+$ are given by density operators of the form $T=2^{-n+1}(pE_++(1-p)E_-)$, where $E_\pm$ is the projection onto $\2H_\pm$. 

\begin{lem}
\label{lem-app-2}
Let $T$ be the density operator for a trace on $\6F^+$, then the two-point function is given by 
\begin{equation}
\tr(T\B(f_1)\B(f_2))=\frac{1}{2}\sprod{\Gamma f_1}{f_2}
\end{equation}
for all $f_1,f_2\in K$. 
\end{lem}
\begin{proof}
We rewrite the density operator $T$ as $T=2^{-n}\I-2^{-n+1}(2p-1)\Theta$ where $\Theta=E_+-E_-$. Let $c_i,c_i^\adj$ be the standard creation and annihilation operators on Fock space and let ${\boldsymbol N}=\sum_jc^\adj_jc_j$ be the particle number operator. There exists an orthonormal basis $(e_I)$ of eigenvectors, i.e., $Ne_I=n_Ie_I$ where there are $n\choose k$ eigenvalues with $n_I=k$. To be more explicit, we choose a orthonormal basis for the one-particle space $(e_i|i=1,\cdots n)$ and associate to each subset $I\subset N:=\{1,\cdots,n\}$ we associate the vector $e_I:=e_{i_1}\wedge e_{i_2}\wedge \cdots e_{i_k}$ with $I=\{i_1,\cdots,i_k\}$ and $i_1<i_2<\cdots<i_k$. Here $\psi\wedge \phi:=2^{-1}(\psi\otimes\phi-\phi\otimes\psi)$ is the antisymmetric tensor product. The standard creation operator acts according to 
\begin{equation}
c_j^\adj e_I:=e_j\wedge e_{i_1}\wedge e_{i_2}\wedge \cdots e_{i_k}
\end{equation}
which vanishes if $j\in I$. If $j\in I$ and let $p(j|I)$ be the number of elements in $I$ that are smaller than $j$, then we get 
\begin{equation}
c_j^\adj e_I=(-1)^{p(j|I)} \delta[j\in N\setminus I]e_{I\cup j} \; .
\end{equation}
where we use the convention $\delta[{\tt true}]=1$ and $\delta[{\tt false}]=0$. Obviously, we get for the annihilation operator:
\begin{equation}
c_j e_I=(-1)^{p(j|I\setminus j)} \delta[j\in I]e_{I\setminus j} \; .
\end{equation}
Indeed, $(e_I)$ is an orthonormal basis of eigenvectors with ${\boldsymbol N}e_I=|I|e_I$. Since $\Theta=(-1)^{\boldsymbol N}$, it follows that $\sprod{e_I}{\Theta c_ic_je_I}=(-1)^{|I|}\sprod{e_I}{c_ic_j e_I}=0$. For the term $\sprod{e_I}{\Theta c_i^\adj c_j e_I}$ we find 
\begin{equation}
\begin{split}
\sprod{e_I}{\Theta c_i^\adj c_j e_I}=(-1)^{|I|}\delta[i\in I]\delta_{ij}  \; .
\end{split}
\end{equation}
Thus the trace of $\Theta c_i^\adj c_j$ can be computed according to 
\begin{equation}
\begin{split}
\tr(\Theta c_i^\adj c_j)=\sum_{I\subset N}(-1)^{|I|}\delta[i\in I]\delta_{ij}
\\
=-\sum_{J\subset N\setminus i}(-1)^{|J|}\delta_{ij}=0 
\end{split}
\end{equation}
Here we have used the fact that $N\setminus i$ contains as many even-elementary subsets as odd-elementary subsets. with an even. This implies $\tr(\Theta\B(f_1)\B(f_2))=0$ and with the help of Lemma~\ref{lem-app-1} we find the desired result: 
\begin{equation}
\tr(T\B(f_1)\B(f_2))=2^{-n}\tr(\B(f_1)\B(f_2))=\frac{1}{2}\sprod{\Gamma f_1}{f_2} \; .
\end{equation}
\end{proof}

With help of the two lemmata above we prove the following theorem:
\begin{thm}
\label{thm-app-1}
There is a unique quasifree normalized trace on the even subalgebra $\6F^+$ which is determined by the covariance $\I/2$.
\end{thm}
\begin{proof}
By Lemma~\ref{lem-app-1}, the quasifree state with covariance $\I/2$ is the normalized trace on the full fermion algebra $\6F$. The restriction to the even subalgebra $\6F^+$ is again a quasifree normalized trace. On the other hand, each state $T$ on $\6F$ that vanishes on the odd part $\6F^-$ and which induces a trace on the even part $\6F^+$ has the form $T=2^{-n}\I+2^{-n+1}(2p-1)\Theta$. If $T$ is quasifree, then it is uniquely determind by the two-point correlation function. Suppose $T=\varrho_S$ is quasifree with covariance $S$, then it follows by Lemma~\ref{lem-app-2} that $S=\I/2$.
\end{proof}
\end{appendix}

\end{document}